\documentstyle[sprocl,epsfig]{article}

\begin{document}

\title{Luminous Cluster Ellipticals as Cosmological Standard Rods?}

\author{Ralf Bender, R.P. Saglia, Bodo Ziegler, Laura Greggio, Ulrich Hopp}

\address{Universit\"ats-Sternwarte, Scheinerstr.\ 1, D--81679
M\"unchen, Germany}

\maketitle\abstracts{We explore the possibility to calibrate massive
cluster ellipticals as cosmological standard rods. The method is based
on the Fundamental Plane relation combined with a correction for
luminosity evolution which is derived from the Mg$-\sigma$
relation. Principle caveats and sources of major 
errors are briefly discussed.\\  We apply the described procedure to
nine elliptical galaxies in two clusters at $z=0.375$ and derive
constraints on the cosmological model.  For the best fitting
$\Lambda$-free cosmological model we obtain: $q_o \approx 0.1$, with
90\% confidence limits being $0 < q_o < 0.7$ (the lower limit being
due to the presence of matter in the Universe). If the inflationary
scenario applies ({\it i.e.}, space has flat geometry), then, for the
best fitting model, matter and $\Lambda$ contribute about equally to
the critical cosmic density (i.e. $\Omega_m \approx \Omega_\Lambda
\approx 0.5$). With 90\% confidence $\Omega_\Lambda$ should be smaller
than 0.9.\\ A more detailed account of the work presented here has 
recently been published~\cite{Betal98}. }

\section{Formation and evolution of elliptical galaxies}

Kinematical studies of nearby masssive elliptical galaxies have shown
that most of these objects have experienced a complex formation
history. Especially kinematically decoupled cores are indicative of
violent similar--mass mergers ~\cite{Bende97}.  While the formation of
field ellipticals may continue until today~\cite{Schwe90}, cluster
ellipticals must have undergone their major merging period at
redshifts $z>0.5$ or even $z>1$.  This follows on the one hand from
the homogenous properties of local cluster
ellipticals~\cite{BLE92,RC93,Colless98} and, on the other hand, from
the very small redshift evolution of their
luminosities~\cite{LTHCF95}, colors~\cite{SED97}, and spectral
indices~\cite{BZB96,ZB97}.  Furthermore, the redshift evolution of the
Fundamental Plane (FP) relation~\cite{D87,DD87,BBF92} of cluster
ellipticals has been found to be consistent with purely passive
evolution of their stellar populations~\cite{KDFIF97}.

Passive evolution of cluster ellipticals up to redshifts of 0.5 to 1
(depending on the cosmological model) is also consistent with
hierarchical galaxy formation models. Even in the standard
$\Omega_o=1$ cold-dark-matter model (with strong low redshift
evolution), neither significant star formation nor accretion processes
take place in cluster ellipticals at $z<0.5$~\cite{K96}.
The star formation activity observed in cluster galaxies at modest
redshifts seems to be confined to blue infalling and/or harrassed disk
galaxies~\cite{MLK97,Bal97} which may turn into intermediate
luminosity S0s today but are unlikely to alter the population of
massive cluster ellipticals at a significant level.

\section{The method }

Our method to calibrate massive cluster ellipticals as cosmological
standard rods relies on the Fundamental Plane (FP) relation combined
with a correction for luminosity evolution. The latter is based on the
distance-independent Mg-$\sigma$ relation and the assumption that
cluster ellipticals evolve passively.

The FP describes the observed tight scaling relation between effective
radius ($R_e$), mean effective surface brightness
($\langle$SB$\rangle_e$) and velocity dispersion ($\sigma$) of cluster
ellipticals: $\log R_e = 1.25\log\sigma + 0.32\langle{\rm SB}\rangle_e
- 8.895$ with $R_e$ in kpc, $\sigma$ in km/s, $\langle$SB$\rangle_e$
in the B-band. The constant -8.895 was derived from the Coma cluster
ellipticals with a Hubble constant of H$_o = 50$km/s/Mpc, the error in
the constant is about 0.01. The FP allows to predict the effective
radii $R_e$ of elliptical galaxies with better than 15\% accuracy from
their velocity dispersions and surface brightnesses, i.e., a distance
dependent quantity ($R_e$) can be estimated from two distance
independent quantities ($\sigma$, $\langle$SB$\rangle_e$).  If a large
enough number of ellipticals are measured per cluster then the
distance to the corresponding cluster can be determined with an
accuracy better than 5\%.

As argued in the previous section, we have strong evidence that
massive ellipticals in rich clusters evolve only passively between now
and at least $z\approx 0.5$.  Passive evolution of the stellar
population can be accurately measured with the Mg$-\sigma$ relation
which is independent of $q_o$ and H$_o$~\cite{BZB96,ZB97}, and is
within current uncertainties the same for clusters of similar
richness/velocity dispersion~\cite{Joerg97}. In local cluster
ellipticals, the Mg$_b$ absorption is tightly coupled to the velocity
dispersion $\sigma$ of the galaxy. Together with the Fundamental Plane
this constrains the scatter in age and metallicity at a given $\sigma$
to be smaller than 15\%~\cite{Colless98}.  Therefore, in the case of
passive evolution, Mg$_b$ decreases with redshift only because the age
of the population decreases.  One can use population synthesis models
to translate the observed ${\rm Mg}_b$ weakening into an estimate of
the B--band luminosity evolution.  Based on recent
models~\cite{Worth94,BC98,BCT96} we obtain consistently $\Delta B =
1.4 \Delta$Mg$_b$ for a Salpeter initial mass function, metallicities
between 1/3 and 3 times solar and ages between 3~Gyr and 15~Gyr.  The
slope of 1.4 shows no dependence (within 0.1) on evolutionary tracks
and other differences in the synthesis models, which demonstrates that
this differential comparison of Mg$_b$ and B-band evolution is quite
robust.

\section{Results from a first application}

As a first application of the method we observed nine elliptical
galaxies in two clusters at a redshift of $z=0.375$. We used the
Hubble Space Telescope to derive effective radii $R_e$ and surface
brightnesses $\langle SB_e\rangle$ and the Calar Alto 3.5m telescope
to obtain velocity dispersions $\sigma$ and
Mg$_b$-indices. Additional photometric calibrations were obtained with
the ESO NTT and the Calar Alto 2.2m telescope. The reduction
procedures are described in more detail in ~\cite{ZB97,Betal98}.

The surface brightness of the $z=0.375$ ellipticals was transformed to
rest--frame B--band, corrected for cosmological dimming ($(1+z)^4$)
and for passive evolution using the ${\rm Mg}_b-\sigma$ relation
described above. For the local comparison sample we used elliptical
and S0 galaxies in the Coma cluster.

Figure~1 (bottom) shows the edge--on view of the Fundamental Plane for Coma
ellipticals (with $\sigma >120$km/s) and for the $z=0.375$
ellipticals.  The angular distances at which a perfect match between
the two samples is achieved are 139~Mpc/h$_{50}$ for Coma and
1400~Mpc/h$_{50}$ for the $z=0.375$ ellipticals (h$_{50}$ is the
Hubble constant in units of 50km/s/Mpc).  The luminosity
correction is calculated based on the {\it mean} offset in the Mg$_b$
absorption of the local and the $z=0.375$ sample, i.e., $\Delta
\langle{\rm SB}\rangle_{B,e} = 1.4 \Delta{\rm Mg}_b =
0.48$mag/arcsec$^2$. We could also have corrected the luminosity
evolution of the objects individually and would have obtained the same
result. In fact, the residuals of the distant ellipticals from the
Fundamental Plane and from the Mg$_b-\sigma$ relation do indeed
correlate with each other in the expected way, though the error bars
are large (Figure~1 top). This supports the idea that the evolution
we see is really due to age (and not metallicity). While for local
samples of ellipticals we cannot conclude reliably that the residuals
from the FP and Mg$_b-\sigma$ relations are correlated and caused by
age~\cite{JFK96}, the effects of age spread must increase with
redshift and may lead to the observed correlation in the residuals.

\begin{figure}
\noindent
\begin{minipage}[b]{0.47\linewidth}
\centering\psfig{figure=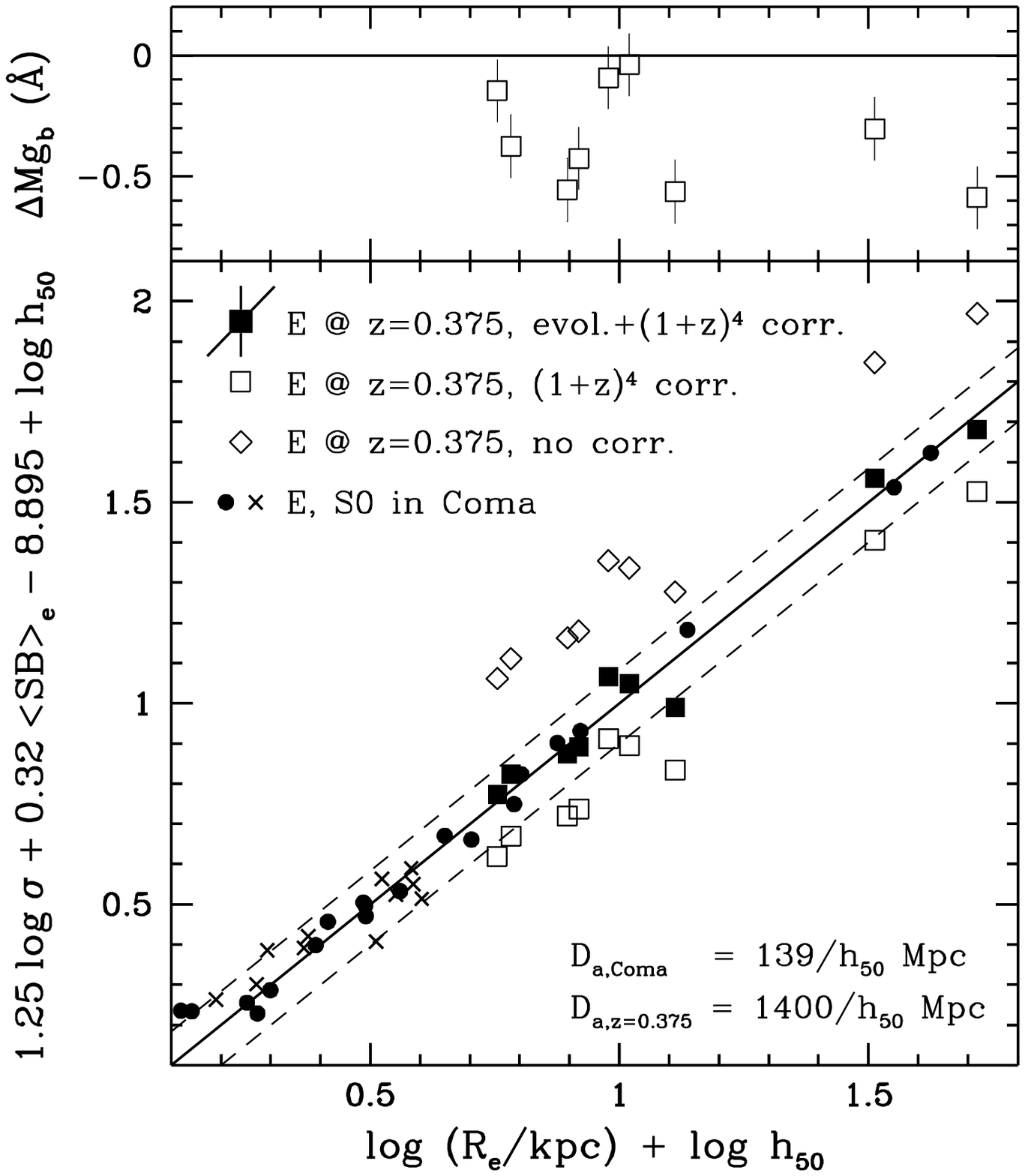,height=7cm}
\caption{{\it Lower panel:} The Fundamental Planes of Coma cluster
ellipticals and evolution-corrected cluster ellipticals at $z=0.375$
in rest frame B-band. The angular distance to $z=0.375$ is derived as
1400~Mpc/h$_{50}$ if Coma is at 139~Mpc/h$_{50}$ distance (h$_{50} =
$H$_o/$(50~km/s/Mpc)).  The upper and lower dashed lines show the FP
relations of the $z=0.375$ objects if their distance were 
1100~Mpc/h$_{50}$ and 1700~Mpc/h$_{50}$, respectively. {\it Upper
panel:} The Mg$_b$ indices of the $z=0.375$ ellipticals relative 
to the Mg$-\sigma$ relation of Coma Es. The luminosity evolution of the
$z=0.375$ Es was derived and corrected using the mean offset of the
Mg$_b$ values. }
\end{minipage}
\hfill
\begin{minipage}[b]{0.47\linewidth}
\centering\psfig{figure=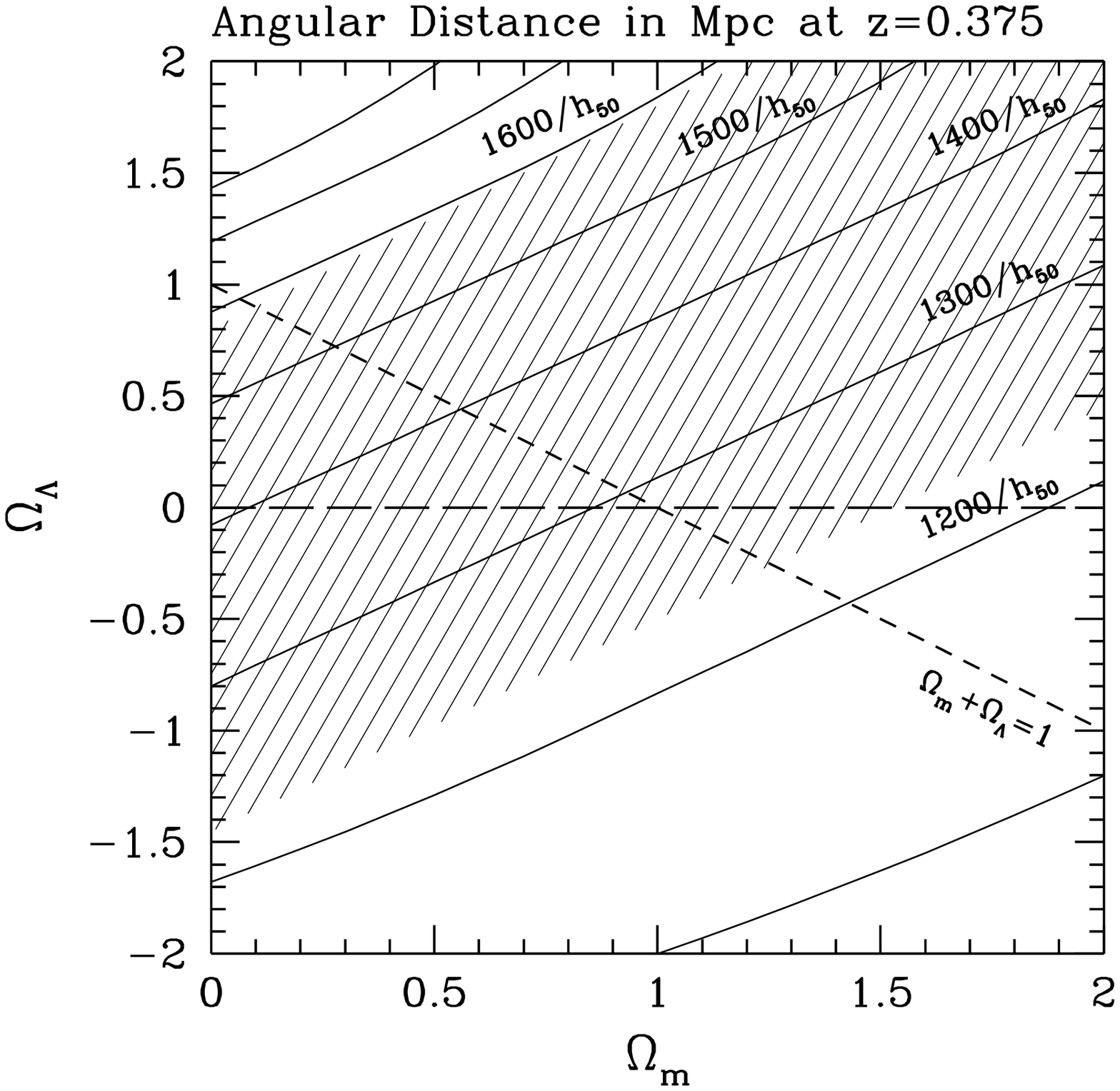,height=7cm}
\caption{The plane of matter density {\it vs.} cosmological constant.
Matter density is given by $\Omega_m$, the cosmological constant is
represented by $\Omega_\Lambda = \Lambda/(3{\rm H}_o^2)$.  Lines of
constant angular distance for a redshift of 0.375 are shown.  The
ellipticals observed at this redshift (Figure 1) allow to constrain
the angular distance to $z=0.375$ which in turn defines a probability
strip in the $\Omega_\Lambda-\Omega_m$ plane. Values for
$\Omega_\Lambda$ and $\Omega_m$ which have $>90$\% likelihood lie in
the shaded area.  The horizontal long--dashed line corresponds to a
Universe without a cosmological constant, the diagonal short--dashed
line to a Universe with no curvature
($\Omega_m+\Omega_\Lambda=1$). }
\end{minipage}
\end{figure}

The cosmological model is constrained on the basis of Figure~1 as
follows. The distance ratio at which the fully corrected FP of the
$z=0.375$ clusters Abell~370 and MS1512+36 matches the FP of the 
Coma cluster is $10.1 \pm 0.8$. With
an angular distance of 139~Mpc/h$_{50}$ to the Coma cluster (corrected to the
Cosmic Microwave Background rest--frame~\cite{FWBDDLT89})
this ratio corresponds to an angular distance of 1400~Mpc/h$_{50}$ to
the $z=0.375$ clusters. For a plausible range of cosmological models, the
distance to the Coma cluster is, because of its proximity, virtually
independent from $q_o$ (at the level of 0.5\%), while the distance to
the $z=0.375$ clusters varies by more than 10\%.  Since the geometry
of the Universe is determined by the ratio of distances, $q_o$ is
independent of the Hubble constant.  The relative error of mean
distance to the $z=0.375$ clusters is about 8\%. The
mean distance to $z=0.375$ and its error give immediate constraints on
the cosmological model (see Figure~2). If the cosmological constant
vanishes then our measurements, together with the observational fact
that there exists matter in the Universe, constrain the cosmological
density parameter to be $0 < \Omega_m < 1.4$, or the cosmological
deceleration parameter to be $0 < q_o < 0.7$, with 90\%
confidence. The best fit suggests $q_o=0.1$. If
the Universe has flat geometry as suggested by inflation, then the
preferred model would have $\Omega_m = \Omega_\Lambda = 0.5$.
$\Omega_\Lambda$ is constrained to fall in the range $-0.25 <
\Omega_\Lambda < +0.9$, again with 90\% confidence.

{\bf Caveats.} To conclude we address some caveats of the proposed
method which can be studied and checked more thoroughly  
once large samples of distant ellipticals become available:\\
(1) Dynamical evolution could alter
the Fundamental Plane relation. However, at present objects of very
different dynamical structure (S0s, rotationally flattened and
anisotropic Es) all lie on the same FP indicating that
it is robust against changes in dynamical structure. \\
(2) Environmental dependences of Fundamental Plane and Mg$-\sigma$
relations. So far no significant dependences have been
found for clusters of similar richness. Moreover, if the
FP varied by even a few percent, peculiar velocities of clusters 
could not be derived~\cite{Burst89,JFK96}.\\
(3) We estimated the combined biasing effect of the sample size, sample
selection and allowed range of the FP parameters by means of Monte
Carlo simulations~\cite{Sagli97}. We find
that the systematic bias of the FP zero-point is still smaller than the
random variation due to sample size.\\
(4) Significant dust absorption can be ruled out, because it
would cause a (B$-$V)--Mg$_b$ relation at $z=0.375$ offset from
the local one, which we did not find.\\
(5) Differential relation between Mg$_b$ weakening and B-band
evolution. We checked various models and found no significant
difference, even for models with different evolutionary tracks. 
The major uncertainty is whether all ellipticals
have a stellar initial mass function with Salpeter slope for masses
around and below the solar mass (for a more detailed discussion
see~\cite{Betal98}).\\
(6) Influence of mixed populations.  We performed tests with varying
fractions of intermediate age populations, all constrained in a way
that the objects at $z=0.375$ would {\it not} be regarded as of $E+A$
type from spectral characteristics~\cite{DG83}.  For plausible
population mixes, the uncertainty in the evolution correction is
significantly smaller than 0.1~mag.

\section*{Acknowledgments}
This work was funded by the Sonderforschungsbereich 375 of the DFG and
by DARA grant 50 OR 9608 5.  RB acknowledges additional support by the
Max--Planck--Gesellschaft.  LG is on leave from Dipartimento di
Astronomia, Universita di Bologna, and was partially supported by the
Alexander-von-Humboldt foundation.

\end{document}